\documentclass{aa}
\usepackage{graphics}

\begin{document}

\thesaurus{08(09.08.1; 02.12.1; 09.09.1 M42; 09.09.1 M43)}

\title{Fluorescence of [\ion{Fe}{ii}] in \ion{H}{ii}~regions
\thanks{Based on observations made with the Isaac Newton Telescope, operated
on the island of La Palma by the Isaac Newton Group in the Spanish
Observatorio del Roque de los Muchachos of the Instituto de Astrof\'\i sica
de Canarias.}}

\author{M. Rodr\'\i guez}

\institute{Instituto de Astrof\'\i sica de Canarias, E-38200 La Laguna, 
Tenerife, Canarias, Spain\\
e-mail: mrodri@ll.iac.es}

\date{Received 13 April 1999; accepted 21 May 1999}

\maketitle
\begin{abstract}

A study of [\ion{Fe}{ii}] lines at various positions within the
\ion{H}{ii}~regions \object{M42} and \object{M43} is presented.
The relative intensities of selected optical [\ion{Fe}{ii}] lines are shown to
be correlated with the intensity of the apparent nebular continuous spectrum.
Since the continuum of \ion{H}{ii}~regions is known to be mostly stellar
radiation scattered by dust intermixed with the emitting gas, these
correlations provide direct evidence for the existence of fluorescent
excitation in the formation process of the [\ion{Fe}{ii}] lines, irrespective
of the prevailing physical state.

\end{abstract}

\section{Introduction}

Recent studies of the [\ion{Fe}{ii}] lines observed in \object{M42} have shown
that some relative intensities between them cannot be accounted for by
assuming that the line excitation is produced only by electron collisions, at
the densities ($N_{\rm e}\la10^4\mbox{ cm}^{-3}$) indicated by the ground
state transitions $^2D-{}^4S$ of [\ion{S}{ii}] and [\ion{O}{ii}]
(e.g.\ Bautista et al.\ \cite{bpp96}; Baldwin et al.\ \cite{bald96}).
In order to solve the ensuing problem, Bautista et al.\ (\cite{bpo94})
postulate that the lines in question have their origin in a high-density
partially ionized layer with $N_{\rm e}\sim10^6\mbox{ cm}^{-3}$
(see also Bautista \& Pradhan \cite{bp95}; Bautista et al.\ \cite{bpp96};
Bautista \& Pradhan \cite{bp98}, BP98 hereafter).
But, alternatively, Lucy (\cite{lucy95}) has shown that the line ratios can be
explained by considering the fluorescent excitation of the \element[+]{Fe}
levels by UV nebular radiation, which is diluted and reprocessed
stellar light, an interpretation further supported by Baldwin et al.\
(\cite{bald96}) and Rodr\'\i guez (\cite{rod96}).
The main arguments presented in all these papers for and against either
interpretation, rest on the comparison of measured [\ion{Fe}{ii}] line
intensity ratios with calculations involving the collision strengths of the
relevant \element[+]{Fe} levels and the spectral distribution of the radiation
field.
Consequently, the reliability of the arguments cannot be readily assessed, and
since not all the calculations use the same set of collision strengths, it is
also difficult to carry out a meaningful comparison of their results.

This paper presents direct observational evidence for the important role
played by fluorescence in the formation of the optical [\ion{Fe}{ii}] lines
in \object{M42} and \object{M43}.
On the basis of this new evidence, the aforementioned interpretations of the
[\ion{Fe}{ii}] line intensities, either in terms of fluorescent excitation or
as a diagnostic for the existence of a high-density partially ionized layer,
will be critically discussed.

\section{The data}

The intensities of line and continuum radiation emitted by various areas
within seven Galactic \ion{H}{ii}~regions, measured by Rodr\'\i guez
(\cite{rod96}), provide the basic data for this discussion.
Because the optical [\ion{Fe}{ii}] lines are intrinsically weak and the
contamination of the true nebular continuum by night-sky light is minimal,
this paper is restricted to the apparently brightest \ion{H}{ii}~regions
\object{M42} and \object{M43}.

The present discussion opens with a comparison between the line and continuum
intensities used in this paper and those available in the literature.
In Table~1 are shown the reddening-corrected intensity ratios,
$I(\lambda)/I(\element{H}\beta)$, of four of the stronger optical
[\ion{Fe}{ii}] lines: $\lambda$4287
($a\>^6D_{9/2}-a\>^6S_{5/2}$), $\lambda$5158+9 ($a\>^4F_{9/2}-a\>^4H_{13/2}$,
$a\>^4F_{7/2}-b\>^4P_{3/2}$), $\lambda$5262 ($a\>^4F_{7/2}-a\>^4H_{11/2}$)
and $\lambda$8617 ($a\>^4F_{9/2}-a\>^4P_{5/2}$).
Besides the uncertainties inherent in the measurement of individual line
intensities (particularly in the fixing of the continuum level), the ratio
$I(8617)/I(\element{H}\beta)$ may be affected by the intensity calibration
between the two different spectral ranges involved, with no lines in common.
This latter uncertainty is estimated to be $\sim15$\%, the degree of
disagreement between the reddening-corrected ratios
$I(\mbox{Pa12})/I(\element{H}\beta)$ and $I(\mbox{Pa13})/I(\element{H}\beta)$
and their recombination values (Hummer \& Storey \cite{hum87}), since
Pa12 and Pa13 are in the same spectral range as $\lambda$8617.
For comparison with some of the
values given in Table~1, the relative [\ion{Fe}{ii}]
line intensities obtained by Osterbrock et al.\ (\cite{otv92}) and Esteban et
al.\ (\cite{est98}) at various positions in \object{M42} are listed in
Table~2.
It  can thus be seen that the ratios of Osterbrock et al.\ (\cite{otv92}) are
quite close to the values for the positions \object{M42}~A--4 and
\object{M42}~A--5 in Table~1.
The values for \object{M42}--1 and \object{M42}--2 in Table~2 are also almost
equal to those for \object{M42}~A--3 and \object{M42}~A--5 in Table~1.
On the whole, then, it would appear that the [\ion{Fe}{ii}] line ratios have a
measurement accuracy of 15\%.

\begin{table}
\caption[ ]{Line and continuum intensities}
\begin{tabular}{llllll}
\hline
\noalign{\smallskip}
\multicolumn{1}{l}{Object$^{\rm a}$} &
	\multicolumn{5}{c}{$[I(\lambda)/I(\element{H}\beta)]\times10^3$} \\
\noalign{\smallskip}
\cline{2-6}
\noalign{\smallskip}
& $\lambda$4287 & $\lambda$5158 & $\lambda$5262 &
	$\lambda$8617 & cont.$^{\rm b}$ \\
\noalign{\smallskip}
\hline
\noalign{\smallskip}
\object{M42} A--1 & 1.63 & 1.34 & 1.00 & 1.04 & 1.9 \\
\object{M42} A--2 & 1.3  & 1.04 & 0.77 & 0.95 & 1.7 \\
\object{M42} A--3 & 0.5  & 0.45 & 0.33 & 0.29 & 2.2 \\
\object{M42} A--4 & 0.79 & 0.85 & 0.59 & 0.59 & 2.0 \\
\object{M42} A--5 & 0.92 & 0.76 & 0.55 & 0.45 & 2.4 \\
\object{M42} A--6 & 0.9  & 0.54 & 0.47 & 0.37 & 3.3 \\
\noalign{\smallskip}
\hline
\noalign{\smallskip}
\object{M42} B--1 & 0.86 & 1.05 & 0.59 & 0.89 & 1.7 \\
\object{M42} B--2 & 0.61 & 0.63 & 0.39 & 0.48 & 2.3 \\
\object{M42} B--3 & 0.87 & 0.60 & 0.32 & 0.58 & 2.1 \\
\object{M42} B--4 & 1.27 & 1.94 & 0.90 & 1.60 & 2.2 \\
\object{M42} B--5 & 1.6  & 1.13 & 0.82 & 0.83 & 3.7 \\
\object{M42} B--6 & 1.3  & 1.01 & 0.54 & 1.06 & 1.9 \\
\noalign{\smallskip}
\hline
\noalign{\smallskip}
\object{M43}--1 & 2.5  & 1.5  & 1.55 & 1.01 & 6.0 \\
\object{M43}--2 & 3.4  & 2.7  & 2.47 & 1.73 & 5.4 \\
\object{M43}--3 & 8.7  & 4.4  & 4.7  & 2.3  & 11.2 \\
\object{M43}--4 & 2.3  & 1.50 & 1.28 & 0.92 & 6.5 \\
\object{M43}--5 & 2.5  & 1.6  & 1.0  & 0.90 & 8.2 \\
\noalign{\smallskip}
\hline
\noalign{\medskip}
\multicolumn{6}{l}{$^{\rm a}$ Slit position A in \object{M42} is centred
	27\arcsec\ to the south of}\\
\multicolumn{6}{l}{\object{$\theta^1$~Ori~C} and orientated east--west;
	slit position B is}\\
\multicolumn{6}{l}{located on the bar. Further information on the areas}\\
\multicolumn{6}{l}{studied is presented in Rodr\'\i guez (\cite{rod96})}\\
\noalign{\smallskip}
\multicolumn{6}{l}{$^{\rm b}$ In units of \AA$^{-1}$. The continuum
	intensities have been}\\
\multicolumn{6}{l}{measured near H$\beta$}\\
\end{tabular}
\end{table}

\begin{table}
\caption[ ]{Previous observations of [\ion{Fe}{ii}] lines}
\begin{tabular}{llllll}
\hline
\noalign{\smallskip}
\multicolumn{1}{l}{Object} &
	\multicolumn{4}{c}{$[I(\lambda)/I(\element{H}\beta)]\times10^3$} &
	 \multicolumn{1}{l}{Ref.$^{\rm a}$} \\
\noalign{\smallskip}
\cline{2-5}
\noalign{\smallskip}
& $\lambda$4287 & $\lambda$5158 & $\lambda$5262 & $\lambda$8617 & \\
\noalign{\smallskip}
\hline
\noalign{\smallskip}
\object{M42}      & 0.86 & 0.87 & 0.54 & 0.67 & (1) \\
\noalign{\smallskip}
\hline
\noalign{\smallskip}
\object{M42}--1 & 0.52  & 0.49 & 0.32 & \dots & (2) \\
\object{M42}--2 & 0.96  & 0.71 & 0.47 & \dots & (2) \\
\noalign{\smallskip}
\hline
\noalign{\medskip}
\multicolumn{6}{l}{$^{\rm a}$ (1) Osterbrock et al.\ (\cite{otv92}); (2)
Esteban et al.\ (\cite{est98})}\\
\end{tabular}
\end{table}

Intensity measurements of the continuous nebular spectrum are not generally
given in papers mainly concerned with the line spectrum, probably because 
they are intrinsically more difficult to carry out, requiring corrections due
to the presence of night-sky light.
The measurements presented here have been made after subtracting separate sky
exposures, the intensities of which had to be
scaled -- by factors between 0.5 and 1.5 -- in order to get the best
cancellation of the sky lines, from the nebular exposures.
In \object{M42} and \object{M43} the night-sky brightness contributes less
than 5\% to the continuum and the sky-subtraction process is thus
inconsequential.
The compilation by Schiffer \& Mathis (\cite{sch74}) of intensity measurements
in the continuous spectrum of \object{M42}, relative to \element{H}$\beta$,
covers the range $1.6\mbox{--}5.5\times10^{-3}\mbox{ \AA}^{-1}$, in agreement
with the measurements presented in Table~1.

\section{Results}

\begin{figure*}
  \resizebox{\hsize}{!}{\includegraphics{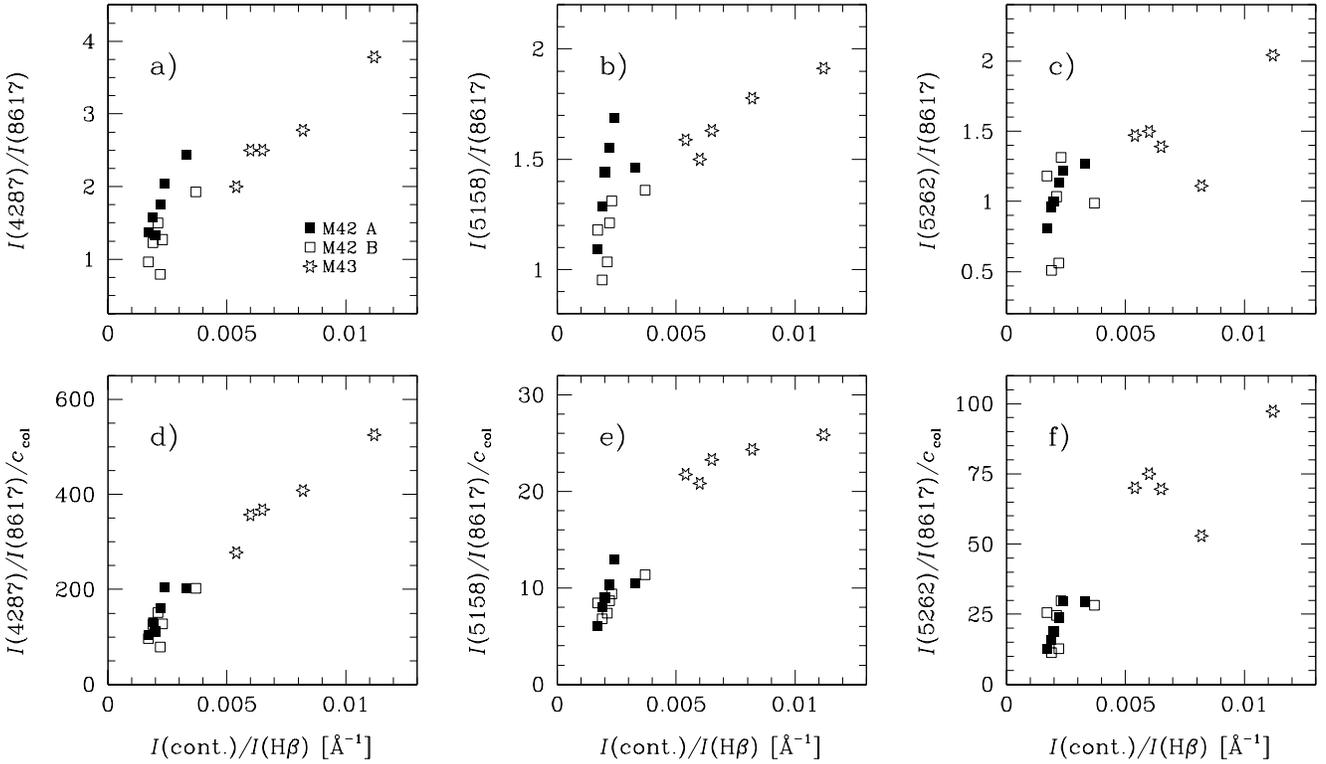}}
    \caption{The $I(4287)/I(8617)$, $I(5158)/I(8617)$ and
    $I(5262)/I(8617)$ ratios by themselves ({\bf a}, {\bf b}, {\bf c}), and
    normalized to their predicted values $c_{\rm col}$ (Bautista \& Pradhan
    \cite{bp96}) for $N_{\rm e}$[\ion{S}{ii}] and $T_{\rm e}$[\ion{N}{ii}]
    ({\bf d}, {\bf e}, {\bf f}), as a function of the nebular continuum
    intensities near H$\beta$ in units of H$\beta$ intensity}
    \label{uno}
\end{figure*}

According to Lucy (\cite{lucy95}) and Baldwin et al.\ (\cite{bald96}), the
line [\ion{Fe}{ii}]~$\lambda$8617 is almost insensitive to the effects
of optical pumping.
The [\ion{Fe}{ii}]~$\lambda$4287 line, on the contrary, is expected to be very
sensitive to fluorescence, since it arises in the  $a\>^6S$ term of the sextet
system to which the $a\>^6D$ ground term also belongs and, therefore,
$a\>^6S$ can be populated by allowed emissions from $z\>^6P^o$ and
$z\>^6D^o$, terms, which in turn are connected to the ground term by allowed
UV transitions.
It follows that if fluorescence plays a role in the formation of
$\lambda$4287, its intensity should be related to that of the radiation field
inducing the fluorescence, which is again in the UV range.
Nevertheless, since the continuous spectrum of \ion{H}{ii} regions is
primarily stellar light scattered by dust coexisting with the emitting gas,
(O'Dell \& Hubbard \cite{oh65}; Peimbert \& Goldsmith \cite{pg72}), its
observed intensity variations in the \element{H}$\beta$ region should
correspond to similar variations in the UV or any other range.
Therefore, the clear correlation shown in Fig.~\ref{uno}a, between
$I(4287)/I(8617)$ and the intensity of the continuum near \element{H}$\beta$,
normalized to the \element{H}$\beta$ intensity, shows indeed that fluorescence
is taking place in the formation process of $\lambda$4287.
The efficiency of fluorescence in enhancing the line intensities, over the
value expected under pure collisional excitation, is shown for $\lambda$4287
in Fig.~\ref{uno}d, which exhibits the correlation between the intensity of
the continuum near \element{H}$\beta$, in units of the \element{H}$\beta$
intensity, and the  $I(4287)/I(8617)$ ratio normalized to the value
$c_{\rm col}$ that the same ratio would take if the line excitation were due
to electron collisions, at densities $N_{\rm e}$[\ion{S}{ii}] and temperatures
$T_{\rm e}$[\ion{N}{ii}] (see Rodr\'\i guez \cite{rod96} for the definitions
of $N_{\rm e}$[\ion{S}{ii}] and $T_{\rm e}$[\ion{N}{ii}]; the values of these
parameters will be given elsewhere).
The values of $I(4287)/I(8617)/c_{\rm col}$ shown in Fig.~\ref{uno}d imply
that fluorescence enhances $\lambda$4287 by two orders of magnitude with
respect to $\lambda$8617.
In this context, it should be noted that the values of $c_{\rm col}$
appearing in Fig.~\ref{uno} were calculated by Bautista \& Pradhan
(\cite{bp96}) using their own collision strengths, although the collision
strengths of Pradhan \& Zhang (\cite{pz93}) and Zhang \& Pradhan (\cite{zp95})
are considered more accurate in BP98.
The calculations of Pradhan \& Zhang do not include the $a\>^6S$ term and
hence cannot be used to calculate the collisional value of $I(4287)/I(8617)$.
But, anyway, the correlation between $I(4287)/I(8617)$ and the intensity in
the continuum is well established for the observed line ratios, i.e.
uncorrected for collisional effects, and can only become tighter when the
line ratios are normalized by $c_{\rm col}$ as calculated with correct values
for the collision strengths.

As far as other [\ion{Fe}{ii}] lines are concerned, only the measured
intensities of $\lambda$5158 and $\lambda$5262 exhibit clear correlations with
the intensity in the continuum, similar although somewhat looser than that
shown by $\lambda$4287, as can be seen in Figs.~\ref{uno}b, \ref{uno}e,
\ref{uno}c and \ref{uno}f.
A fluorescent contribution to other lines, however, cannot be excluded, as
for example, $\lambda$4244+5 ($a\>^4G_{7/2}-a\>^4F_{7/2}$,
$a\>^4G_{11/2}-a\>^4F_{9/2}$), $\lambda$4277
($a\>^4F_{7/2}-a\>^4G_{9/2}$) and $\lambda$5334
($a\>^4F_{5/2}-a\>^4H_{9/2}$), are weaker than the transitions
shown in Fig.~\ref{uno}, and therefore could be measured only in a few
positions in \object{M42}, insufficient to provide sets of data with definite
trends.
The  $\lambda$4815 ($a\>^4F_{9/2}-b\>^4F_{9/2}$) and
$\lambda$7155 ($a\>^4F_{9/2}-a\>^2G_{9/2}$) lines are as strong as those
of Fig.~\ref{uno}, but $\lambda$4815 is blended with
\ion{Si}{iii}~$\lambda$4813 and \ion{S}{ii}~$\lambda$4815 (Esteban et al.\
\cite{est98}), and $\lambda$7155 is insensitive to fluorescence, since
it arises in the doublet system.

The fluorescence effects clearly illustrated in Fig.~\ref{uno}, especially
for the lines $\lambda$4287 and $\lambda$5158, give grounds for considering
 that
most [\ion{Fe}{ii}] lines may be significantly affected by radiative
excitation, depending in a very complicated fashion on the structure of the
\element[+]{Fe} ion.
The observed strength of the [\ion{Fe}{ii}] lines can therefore  be explained
in terms of a line formation process based on physical principles applicable
under the conditions characteristic of conventional nebular models (density,
temperature and state of ionization).
Since fluorescent excitation of [\ion{Fe}{ii}] is not important for densities
greater than $10^5\mbox{ cm}^{-3}$ (BP98), it would appear that the use of the
[\ion{Fe}{ii}] lines as a diagnostic for the existence of a high-density layer
is completely inappropriate.
The proponents of this high-density model have attempted to use the intensity
ratio $I(\lambda6300+\lambda6363)/I(\lambda5577)$ of nebular [\ion{O}{i}]
lines as independent evidence for the high-density layer in their model
(Bautista \& Pradhan \cite{bp95}; BP98) , but the
nebular [\ion{O}{i}] lines are difficult to measure accurately because of
their contamination by the strong night-sky emission in the same
[\ion{O}{i}] lines, especially the auroral feature at $\lambda$5577.
The ratios obtained from recent and reliable measurements of the nebular
component of this line in \object{M42} (Esteban et al.\ \cite{est99}; see also
Baldwin et al.\ \cite{bald96}), have shown that the [\ion{O}{i}] line ratio is
in fact quite consistent with the line formation taking place at moderate
densities.
Besides, recent measurements of [\ion{Fe}{ii}] lines insensitive to
fluorescence in the 1--2~$\mu$m spectrum of the ``bar'' in \object{M42}
(Marconi et al.\ \cite{mar98}; Luhman et al.\ \cite{luh98}) indicate that the
densities at their levels of formation are in the range
$10^3\mbox{--}10^4\mbox{ cm}^{-3}$.

Independently of their argument based on the nebular [\ion{O}{i}] lines, BP98
also suggest that several measured [\ion{Fe}{ii}]
line ratios, when compared with model predictions, imply the existence of a
high-density emitting layer, even allowing for the contribution of lower
density layers to the line intensities.
However, when comparing the predictions of the two [\ion{Fe}{ii}] line
formation models under discussion, it should be kept in mind that both rely on
calculated collision strengths for the lines, which must be used with
caution.
An example of the uncertainties affecting the collisions strengths 
well illustrates the problem.
The calculations of Pradhan \& Zhang (\cite{pz93}) and Zhang \& Pradhan
(\cite{zp95}) are considered in BP98 to have relatively low uncertainties, but
they do not consider the  $\lambda$4287 and $\lambda$7155 transitions.
The two available sets of collision strengths dealing with $\lambda$4287
(Bautista \& Pradhan \cite{bp96}; BP98) lead to predicted values for the
$I(4287)/I(8617)$ ratio that differ by a factor of 10 for any density value.
Therefore the values derived for $I(4287)/I(8617)$ and $I(7155)/I(8617)$ may
be quite uncertain, and these ratios are precisely those implying more clearly
the existence of high-density emitting regions according to BP98.

In conclusion, it can be said that there is no compelling evidence for the
presence of high-density regions to explain the origin of the [\ion{Fe}{ii}]
lines.
Those in the near infrared must arise in regions of moderate density, while
the optical lines have been clearly shown here to be affected by
fluorescence, of significance only at moderate densities.

\subsection{On the efficiency of fluorescent excitation}

It has been argued by BP98 that photoexcitation of [\ion{Fe}{ii}] lines is a
relatively inefficient mechanism, since the ground state of the
\element[+]{Fe} ion is $a\ {}^6D_{9/2}$
whereas most of the observed lines arise in the quartet system.
According to BP98, photoexcitation of these quartet levels must occur through
intercombination transitions, with transition probabilities much lower
than those of permitted transitions.
However, even at moderate densities ($N_{\rm e}\sim10^3\mbox{ cm}^{-3}$)
the lowest level of the  $a\ {}^4F_{9/2}$ quartets has an appreciable
population (Osterbrock et al.\ \cite{otv92}), and, therefore, higher quartet
levels can be populated through permitted transitions from this level.

The absence of some \ion{Fe}{ii} lines in the spectra of \object{M42} has been
considered by BP98 as further evidence against fluorescent excitation.
In particular, \ion{Fe}{ii}~$\lambda$5169
($z\>^6P^o_{7/2}-a\>^6S_{5/2}$) is mentioned as the main
transition that would contribute to populate $a\>^6S_{5/2}$ radiatively.
The intensity of $\lambda$5169 should then be about 70\% that of
$\lambda$4287, according to BP98.
Since the upper limit of the relative intensities of these lines has been
estimated to be  0.1 for \object{M42}, BP98 conclude that less that 20\% of
the [\ion{Fe}{ii}]~$\lambda$4287 intensity can be explained by fluorescent
excitation.

The spectra available for \object{M42} and \object{M43} show a weak feature at
$\lambda\sim5169~\mbox{\AA}$ whose intensity is about 10\% that of
$\lambda$4287, in accord with the upper limit mentioned by BP98.
In view of the clear demonstration in Fig.~\ref{uno} of the importance of
fluorescence effects in the formation of $\lambda$4287, this result is
puzzling and difficult to explain, as it is also the extremely low
contribution of fluorescence to the ratio $I(4287)/I(8616)$ estimated by BP98
(see their Fig.~4.6d).
One way out of this difficulty would be to consider as alternative radiative
excitation mechanisms of the level $a\ {}^6S_{5/2}$, transitions to levels
$y\>^6P^o$ or $x\>^6P^o$ (with energies 0.57 and 0.72~Ry, respectively),
implying the absorption of photons with
$\lambda=1608\mbox{ or }1261~\mbox{\AA}$.
These sextets are comparable in energy with some of the quartets that BP98
consider when calculating the fluorescence effects on [\ion{Fe}{ii}] emission,
but they are not included in the set of collision strengths used by BP98, and
therefore, these sextets are not considered in their calculations.

The effects of fluorescent excitation on the [\ion{Fe}{ii}] line intensities
in \object{M42} have also been calculated by Baldwin et al.\ (\cite{bald96}).
Unfortunately, neither the $a\ {}^6S$ nor the 
$a\ {}^2G$ terms are included in
the set of collision strengths they use (Pradhan \& Zhang \cite{pz93};
Zhang \& Pradhan \cite{zp95}), and lines like $\lambda$4287 and $\lambda$7155
are not considered in their calculations.
The same collision strengths are used by BP98 for the lines in common, but
their predicted line ratios are somewhat different from those calculated by
Baldwin et al.\ (\cite{bald96}).
Nevertheless, the latter authors conclude that fluorescent excitation in a
region of moderate density can explain the [\ion{Fe}{ii}] spectrum observed by
Osterbrock et al.\ (\cite{otv92}), while opposite conclusions are advanced by
BP98.
The different approaches to the problem of both papers make it difficult to
find the reasons for the discrepancies.
The differences in the contribution of fluorescence to the line ratios
presented by Baldwin et al.\ (\cite{bald96}) and BP98 can thus be considered
to reflect the uncertainties involved in the calculation of fluorescence
effects in a complex ion like \element[+]{Fe}.

In summary, none of the available calculations faithfully reproduces  the
observed [\ion{Fe}{ii}] spectra, but it should be borne in mind that the
effects of UV pumping on the [\ion{Fe}{ii}] line ratios can be quite
different from those calculated so far, since the contribution to the pumping
of the dust-scattered light -- whose relative intensity increases with
frequency -- has not yet been taken into account.
The change in the spectral distribution of the diffuse radiation field would
imply that terms like $z\>^4G^o$ (located 0.55~Ry above the ground level) and
the sextets mentioned above ($y\>^6P^o$ and $x\>^6P^o$) would have greater
contributions to the pumping, thereby increasing the fluorescence effects on
lines like $\lambda$4287, $\lambda$4815, $\lambda$5158, $\lambda$5262 or
$\lambda$5334.

\section{Conclusions}

The relative intensities of the [\ion{Fe}{ii}] lines in the infrared spectra
of \object{M42} imply densities in the range
$10^3\mbox{--}10^4\mbox{ cm}^{-3}$ (Marconi et al.\ \cite{mar98}; Luhman et
al.\ \cite{luh98}), but the optical [\ion{Fe}{ii}] spectrum
cannot be reproduced assuming pure collisional excitation at these low
densities, independently of the set of collision strengths used in the
calculations.
Two additional agents for the excitation of the upper levels of the optical
lines have been proposed: UV pumping (Lucy \cite{lucy95}) and emission at very
high densities $N_{\rm e}\sim10^6\mbox{ cm}^{-3}$
(Bautista et al.\ \cite{bpo94}).
The available calculations based on these two processes (Baldwin et al.\
\cite{bald96}; BP98) encounter certain difficulties
when trying to reproduce faithfully the observed [\ion{Fe}{ii}] line ratios.
However, these calculations depend on the values used for the collision
strengths, which have an accuracy that it is 
difficult to estimate, on the completeness
of the set of levels considered in the pumping processes and on the spectral
intensity distribution of the radiation field involved.
Consequently, the overall reliability of the results is difficult to assess.

The observations presented here have been shown to imply the importance of
fluorescence processes on the formation of the optical [\ion{Fe}{ii}]
emission.
This conclusion is independent of any calculation and renders  the
assumption of a high-density emitting layer unnecessary.
Further implications are the unreliability of the available collision
strengths for \element[+]{Fe} (at least for some sextets and the doublets),
and the need for further calculations on fluorescence that take into account
the contribution of dust-scattered light to the radiation field.

\begin{acknowledgements}
I am very grateful to Guido M\"unch and Antonio Mampaso for their advice
during the development of this project and their contribution to the
improvement of this manuscript. I also thank Terry Mahoney for revising the
English text.
\end{acknowledgements}



\begin{thebibliography}{}
\bibitem[1996]{bald96}
Baldwin J.A., Crotts A., Dufour R.J., et al., 1996, ApJ 468, L115
\bibitem[1995]{bp95}
Bautista M.A., Pradhan A.K., 1995, ApJ 442, L65
\bibitem[1996]{bp96}
Bautista M.A., Pradhan A.K., 1996, A\&AS 115, 551
\bibitem[1998]{bp98}
Bautista M.A., Pradhan A.K., 1998, ApJ 492, 650 (BP98)
\bibitem[1994]{bpo94}
Bautista M.A., Pradhan A.K., Osterbrock D.E., 1994, ApJ 432, L135
\bibitem[1996]{bpp96}
Bautista M.A., Peng J., Pradhan A.K., 1996, ApJ 460, 372
\bibitem[1998]{est98}
Esteban C., Peimbert M., Torres-Peimbert S., Escalante V., 1998, MNRAS
295, 401
\bibitem[1999]{est99}
Esteban C., Peimbert M., Torres-Peimbert S., 1999, A\&A 342, L37 
\bibitem[1987]{hum87}
Hummer D.G., Storey P.J., 1987, MNRAS 224, 801
\bibitem[1995]{lucy95}
Lucy L.B., 1995, A\&A 295, 555
\bibitem[1998]{luh98}
Luhman K.L., Engelbracht C.W., Luhman M.L., 1998, ApJ 499, 799
\bibitem[1998]{mar98}
Marconi A., Testi L., Natta A., Walmsley C.M., 1998, A\&A 330, 696
\bibitem[1965]{oh65}
O'Dell C.R., Hubbard W.D. 1965, ApJ 142, 591
\bibitem[1992]{otv92}
Osterbrock D.E., Tran H.D., Veilleux S., 1992, ApJ 389, 305
\bibitem[1972]{pg72}
Peimbert M., Goldsmith D.W. 1972, A\&A 19, 398
\bibitem[1993]{pz93}
Pradhan A.K., Zhang H.L., 1993, ApJ 409, L77
\bibitem[1996]{rod96}
Rodr\'\i guez M., 1996, A\&A 313, L5
\bibitem[1974]{sch74}
Schiffer III F.H., Mathis J.S., 1974, ApJ 194, 597
\bibitem[1995]{zp95}
Zhang H.L., Pradhan A.K., 1995, A\&A 293, 953
\end{thebibliography}
\end{document}